\documentclass[conference,10pt]{IEEEtran}

\pdfoutput=1
\usepackage{cite}
\usepackage{authblk}
\usepackage{caption}
\usepackage[switch]{lineno}
\setpagewiselinenumbers

\usepackage{enumitem}
\usepackage{amsfonts}
\usepackage{amsmath}
\usepackage{amssymb}
\usepackage{color}
\usepackage[dvipsnames]{xcolor}
\usepackage{colortbl}
\definecolor{darkblue}{rgb}{0.1,0.2,0.6}
\definecolor{darkred}{rgb}{0.8,0.1,0.2}
\definecolor{Gray}{gray}{0.9}
\usepackage[colorlinks,citecolor=darkblue,linkcolor=darkred,urlcolor=darkblue]{hyperref}	
\usepackage[all]{hypcap} 

\usepackage{graphicx}
\graphicspath{{images/}}

\usepackage{enumitem}
\usepackage{listings}
\usepackage[group-separator={,}]{siunitx}

\definecolor{codegreen}{rgb}{0,0.6,0}
\definecolor{codegray}{rgb}{0.5,0.5,0.5}
\definecolor{codepurple}{rgb}{0.58,0,0.82}
\definecolor{backcolour}{rgb}{0.98,0.98,0.96}

\lstdefinestyle{mystyle}{
    backgroundcolor=\color{backcolour},   
    commentstyle=\color{codegreen},
    keywordstyle=\color{magenta},
    numberstyle=\tiny\color{codegray},
    stringstyle=\color{codepurple},
    basicstyle=\ttfamily\footnotesize,
    breakatwhitespace=false,         
    breaklines=true,                 
    captionpos=b,                    
    keepspaces=true,                 
    showspaces=false,                
    showstringspaces=false,
    showtabs=false,                  
    tabsize=2
}

\usepackage{xparse} 
\usepackage{adjustbox}
\ExplSyntaxOn 
\NewDocumentCommand{\cinline}{v} {{\small\adjustbox{bgcolor=backcolour}{\texttt{#1}}}}
\ExplSyntaxOff

\def\quail{
	Quantum Artificial Intelligence Lab. (QuAIL), 
    NASA Ames Research Center, Moffett Field, CA 94035, USA
}

\def\usra{
	USRA Research Institute for Advanced Computer Science (RIACS),
	615 National Ave., Mountain View, California 94043, USA
}

\def\kbr{
  KBR, Inc., 601 Jefferson St., Houston, TX 77002, USA
}

\def\HybridQ{\textbf{\texttt{HybridQ}}}

\begin{document}

\title{HybridQ: A Hybrid Simulator for Quantum Circuits}

\author[1,2,*]{Salvatore Mandr\`a}
\author[1,3]{Jeffrey Marshall}
\author[1]{Eleanor G. Rieffel}
\author[1]{Rupak Biswas}
\affil[1]{\small\quail}
\affil[2]{\kbr}
\affil[3]{\usra}
\affil[*]{\href{mailto:salvatore.mandra@nasa.gov}{salvatore.mandra@nasa.gov}}

\maketitle

\begin{abstract}
  Developing state-of-the-art classical simulators of quantum circuits is of
  utmost importance to test and evaluate early quantum technology and
  understand the true potential of full-blown error-corrected quantum
  computers.  In the past few years, multiple theoretical and numerical
  advances have continuously pushed the boundary of what is classically
  simulable, hence the development of a plethora of tools which are often
  limited to a specific purpose or designed for a particular hardware (e.g.
  CPUs vs. GPUs).  Moreover, such tools are typically developed using tailored
  languages and syntax, which makes it hard to compare results from, and create
  hybrid approaches using, different simulation techniques.
  To support unified and optimized use of these techniques across platforms, we
  developed \HybridQ, a highly extensible platform designed to provide a common
  framework to integrate multiple state-of-the-art techniques to run on a
  variety of hardware. The philosophy behind its development has been driven by
  three main pillars: \emph{Easy to Use}, \emph{Easy to Extend}, and \emph{Use
  the Best Available Technology}. The powerful tools of \HybridQ\ allow users
  to manipulate, develop, and extend noiseless and noisy circuits for different
  hardware architectures. \HybridQ\ supports large-scale high-performance
  computing (HPC) simulations, automatically balancing  workload among
  different processor nodes and enabling the use of multiple backends to
  maximize parallel efficiency.  Everything is then glued together by a simple
  and expressive language that allows seamless switching from one technique to
  another as well as from one hardware to the next, without the need to write
  lengthy translations, thus greatly simplifying the development of new hybrid
  algorithms and techniques.
\end{abstract}

\section{Introduction and Motivations}

Since the historical milestone of quantum supremacy achieved by the Google team
\cite{q_suprem_google}, quantum technology has continued to advance at an
incredibly fast pace. More and more laboratories and companies have developed
their own quantum hardware using a variety of materials, including
superconducting qubits, trapped-ions, or photonic quantum chips paving the road
for full-blown error-corrected quantum computers. At the same time, multiple
theoretical and numerical advances have raised the bar for a potential quantum
advantage and pushed the boundaries of what is classically simulable
\cite{mi2021information, huang2020classical}. Among the different techniques to
simulate large-scale quantum circuits, three are widely used: state vector
simulation \cite{guerreschi2020intel, pednault2019leveraging}, tensor
contraction \cite{villalonga2020establishing, villalonga2016flexible,
gray2021hyper, zlokapa2020boundaries, pan2020contracting, markov_tn,
markov2018quantum}, and expansion around efficiently simulable quantum circuits
\cite{aaronson_gottesmann_stabilizer_sim, Bravyi2019simulationofquantum,
bravyi2016improved, jozsa2008matchgates, aharonov2003simple}. 
To aid the development of new quantum algorithms and the large-scale simulation
of quantum circuits, a plethora of tools have been developed.  Many of these
tools, however, consider only a few simulation techniques and can typically be
run only on specific hardware (e.g. CPUs or GPUs). Moreover, such tools use
tailored languages and syntax specific to an application, making it hard to
compare results and develop hybrid algorithms.  
To support unified and optimized use of these techniques across platforms, we
developed \HybridQ\, a highly extensible platform designed to provide a
framework to integrate multiple state-of-the-art simulation techniques, and run
seamlessly on a variety of hardware types.
\HybridQ\ is entirely written in Python, and only uses compiled languages such
as $\text{C}^{++}$ to achieve state-of-the-art performance in core parts of the
simulation. The simple and expressive language developed in \HybridQ\ allows
users to design and manipulate both noiseless and noisy circuits, and to
seamlessly switch from one simulation technique to another without the burden
of rewriting an entirely new algorithm. \HybridQ\ also allows simulation on
multiple backends, including CPUs, GPUs, and TPUs, both on single nodes or on
heterogeneous HPC clusters.\\

Among the many innovations introduced in \HybridQ, it is worth highlighting:
\begin{itemize}
    \setlength\itemsep{1em}
    \item Fully integrates multiple techniques under the same common framework:
    \begin{itemize}
        \item \textbf{State Vector Simulation}: A multi-threaded $\text{C}^{++}$ core
          which uses AVX instructions to achieve state-of-the-art performance
          in matrix-vector multiplication of quantum states. The $\text{C}^{++}$ core,
          which uses a similar syntax of \texttt{numpy.dot}, is integrated into
          \HybridQ\ whenever a matrix-vector multiplication is needed.
        
        \item \textbf{Tensor Contraction}: \HybridQ\ fully integrates
          \texttt{cotengra}\cite{cotengra} (one of the most advanced tools used
          to identify optimal contractions of tensors) and
          \texttt{quimb}\cite{quimb} to run large-scale simulations of quantum
          circuits.
        
        \item \textbf{Clifford Expansion}: A novel approach based on fusing
          together multiple non-Clifford gates and then finding the smallest
          number of stabilizer operators to represent the fused result was
          devised. Further, a parallelized branch-and-bound technique was
          introduced to minimize the number of branches, thereby greatly
          reducing the computational cost of the simulation.
    \end{itemize}

    \item \HybridQ\ supports large-scale simulations on heterogeneous HPC
      clusters via MPI.  More precisely, without any prior knowledge from the
      user, it automatically splits and balances the workload among different
      processor nodes and allows the use of multiple backends to maximize
      parallel efficiency.

    \item Simulations of both noiseless and noisy circuits are supported in
      \HybridQ. More importantly, it fully integrates noise in its framework,
      allowing the simulation of noisy circuits regardless of the numerical
      technique chosen (including tensor contraction).

    \item Instead of developing monolithic gates which are hard to extend and
      develop, the concept of ``properties'' is introduced to speedup and
      simplify the development of new gates. Each property is developed
      independently and enables new abstract ``features'' (such as the concept
      of having qubits, of having parameters, or simply of being unitary).
      Gates are then built by collecting multiple properties. In this manner,
      any improvement on a specific feature will immediately percolate to any
      gate using such a property. This approach avoids multiple development of
      the same features on different gates, maximizing code recycling.

    \item \HybridQ\ fully integrates the use of ``symbols'' (as in
      \texttt{sympy}\cite{sympy}) in almost every part of the framework. For
      instance, one may use \texttt{sympy.symbols} to parametrize gates, which
      can be eventually specified at a later time.
\end{itemize}

\section{Gates and Properties}

Gates are at the core of any simulator of quantum circuits, as they represent
the basic operations to manipulate quantum states. Therefore, gates must be
expressive enough to allow arbitrary simulations but, at the same time, simple
enough to allow users of any level to intuitively simulate existing quantum
hardware.

To this end, \HybridQ\ uses a bottom-up approach based on the development of
``properties''. More precisely, instead of building monolithic gates, each
one with its own specification and design, properties are build independently so
that multiple gates can inherit from them. This not only allows a high reuse
of existing code, but it drastically simplify the development and extension of
existing gates as new properties are created and deployed. Indeed, basic
properties used across the library are defined in
\cinline{base.property}, and more specialized properties are built on
top of them. As an example, the property to have ``qubits'' can be
implemented as simply as follows:
\begin{lstlisting}[language=Python]
from hybridq import base

@base.staticvars('qubits')
class Qubits(base.__Base__):
    """
    Class with 'qubits'.
    """
    pass

# Generate new type
QubitGate = base.generate('QubitGate', 
                          (Qubits, ), 
                          qubits=(1, 2, 3))


# Generate new gate
gate = QubitGate()

# Output gate qubits
gate.qubits
> (1, 2, 3)
\end{lstlisting}

In \HybridQ, there are two types of gates: gates acting on quantum states
(\cinline{gate.Gate}) and gates acting on density matrices
(\cinline{dm.gate.Gate}). In both cases, gates can be specified by their common
names. For instance:
\begin{lstlisting}[language=Python,mathescape=True]
from hybridq.gate import Gate
from sympy.abc import g
import numpy as np

# Generate a Hadamard gate
Gate('H').on([1])**1.2
> Gate_H(name='H', qubits=(1,), M=numpy.ndarray(shape=(2, 2), dtype=float64))**1.2

# Generate an X-Rotation
Gate('RX', params=[np.pi])**2.2
> Gate_RX(name='RX', n_qubits=1, $\phi$=2.2$\pi$)

# Generate a controlled-phase gate
Gate('CPHASE', qubits=['a', 'b'], params=[g]).adj()
> Gate_CPHASE^+(name='CPHASE', qubits=('a', 'b'), params=(g,))
\end{lstlisting}
allows to specify the Hadamard gate and the controlled-phase gate respectively.
Gates can also be specified directly providing a matrix:
\begin{lstlisting}[language=Python]
from hybridq.gate import MatrixGate
import numpy as np

# Generate a gate by specifying its matrix
MatrixGate(np.eye(2), qubits=['qubit'])**2.2
> MatrixGate(name='MATRIX', 
>            qubits=('qubit',), 
>            M=numpy.ndarray(shape=(2, 2), 
>            dtype=float64))**2.2
\end{lstlisting}
If a closer look is taken at the specific inheritance, one can see that
different gates may inherit from different properties to better describe them:
\begin{lstlisting}[language=Python]
from hybridq.gate import Gate

# Output inheritance for Hadamard and X-Rotation
type(Gate('H')).mro()
> [hybridq.base.base.Gate_H,
>  hybridq.gate.gate.BaseGate,
>  hybridq.gate.property.CliffordGate,
>  hybridq.gate.property.MatrixGate,
>  hybridq.gate.property.SelfAdjointUnitaryGate,
>  hybridq.gate.property.UnitaryGate,
>  hybridq.gate.property.PowerMatrixGate,
>  hybridq.gate.property.PowerGate,
>  hybridq.gate.property.QubitGate,
>  hybridq.base.property.Tags,
>  hybridq.base.property.Name,
>  hybridq.base.base.__Base__,
>  object]

type(Gate('RX')).mro()
> [hybridq.base.base.Gate_RX,
>  hybridq.gate.gate.BaseGate,
>  hybridq.gate.property.RotationGate,
>  hybridq.gate.property.ParamGate,
>  hybridq.base.property.Params,
>  hybridq.gate.property.UnitaryGate,
>  hybridq.gate.property.PowerMatrixGate,
>  hybridq.gate.property.PowerGate,
>  hybridq.gate.property.QubitGate,
>  hybridq.base.property.Tags,
>  hybridq.base.property.Name,
>  hybridq.base.base.__Base__,
>  object]
\end{lstlisting}

Beyond the most common gates, \HybridQ\ provides a variety of useful
\cinline{gate.Gate} that allows great freedom in specifying quantum circuits:
\begin{itemize}
  \item \cinline{StochasticGate(gates, p)}: \cinline{gates} are
    sampled with probability \cinline{p} every time the gate is applied to a
    quantum state,
  \item \cinline{TupleGate(gates)}: \cinline{gates} are gathered
    together and applied to the quantum state one by one,
  \item \cinline{Control(c_qubits, gate)}: The application of
    \cinline{gate} to the quantum state is controlled by \cinline{c_qubits},
  \item \cinline{Projection(state, qubits)}: The quantum state is
    projected to \cinline{state},
  \item \cinline{Measure(qubits)}: A measurement operation on
    \cinline{qubits} is applied to the quantum state,
  \item \cinline{FunctionalGate(f, qubits)}: \cinline{f} is applied
    to the quantum state. Observe that \cinline{f} can be an \emph{arbitrary}
    function, allowing the easy and straightforward implementation of oracles such
    as the Grover gate \cite{grover1996fast} or QAOA simulations \cite{farhi2016quantum, wang2018quantum, hadfield2019quantum},
  \item \cinline{SchmidtGate(gates, s)}: This gate is used to
    represent decomposed gates in the form $\sum_{ij} s_{ij} L_i \otimes R_j$,
    with $L_i$ and $R_j$ being arbitrary gates. Arbitrary gates can be
    decomposed in \cinline{SchmidGate} by using
    \cinline{gate.utils.decompose}. Similarly, \cinline{SchmidGate} can
    be merged back to a \cinline{MatrixGate} by using
    \cinline{gate.utils.merge}.
\end{itemize}
  
At the moment, \HybridQ\ implements the following \cinline{dm.gate.Gate} for
density matrix simulations:
\begin{itemize}
  \item \cinline{KrausSuperGate(gates, s)}: The \cinline{KrausSuperGate} is a
    specialization of the \cinline{SchmidtGate} for density matrices. More
    specifically, the gate implements:
    $$
    \rho \rightarrow \mathcal{E}(\rho) = \sum_{ij} s_{ij} L_i \rho R_j^\dagger,
    $$
    with $\rho$ being the density matrix, and $L_i$ and $R_j$ being arbitrary
    gates.
  \item \cinline{MatrixSuperGate}: This gate is the operator representation of
    the superoperator defining a quantum map. More precisely, once $\rho$ is
    vectorized, \cinline{MatrixSuperGate} is applied to $\rho$ using a
    matrix-vector multiplication.
\end{itemize}

All \cinline{gate.Gate} and \cinline{dm.gate.Gate} provide a useful way
to ``tag'' them. Tagging can be used to identify and/or filter specific gates:
\begin{lstlisting}[language=Python]
from hybridq.gate.utils import get_available_gates
from hybridq.gate import Gate
import numpy as np

# Generate random gates with tags
gates = [
    Gate(np.random.choice(get_available_gates()),
         tags=dict(a=np.random.randint(2)))
           for _ in range(10)
]

# Filter only gates with tags['a'] == 0
list(filter(lambda g: g.tags['a'] == 0, gates))
> [Gate_U3(name='U3', 
>          n_qubits=1, 
>          n_params=3, 
>          tags={'a': 0}),
>  Gate_CX(name='CX', 
>          n_qubits=2, 
>          M=numpy.ndarray(shape=(4, 4), 
>                          dtype=int64), 
>          tags={'a': 0}),
>  Gate_ISWAP(name='ISWAP', 
>             n_qubits=2, 
>             M=numpy.ndarray(shape=(4, 4),
>                             dtype=complex128), 
>             tags={'a': 0}),
>  Gate_FSIM(name='FSIM', 
>            n_qubits=2, 
>            n_params=2, 
>            tags={'a': 0})]
\end{lstlisting}

\subsection{Noisy gates \label{sec:noisy_gates}}

\HybridQ\ fully supports multiple noise channels (\cinline{noise.channel}) which are different
specializations of the more general \cinline{KrausSuperGate}.
Among them, it is worth mentioning:
\begin{itemize}
  \item \cinline{GlobalPauliChannel}: A general channel describing the map $\rho
    \rightarrow \sum_{\vec{i}, \vec{j}} s_{\vec{i}, \vec{j}} \sigma_{\vec{i}} \rho
    \sigma_{\vec{j}}$, where $\vec{i}, \vec{j}$ are length $n$ vectors (for $n$ qubits)
    with entries in $\{0,1,2,3\}$ describing a tensor product of single-qubit
    Pauli operators.
  \item \cinline{GlobalDepolarizingChannel}: A special case of
    \cinline{GlobalPauliChannel}, describing $\rho \rightarrow (1-p)\rho +
    \frac{p}{d}\mathbb{I}$, where $d=2^n$ is the dimension, $\mathbb{I}$ the
    identity matrix, and $p\in[0,1]$ a probability. 
  \item \cinline{DephasingChannel}: A single qubit dephasing channel of the
    form $\rho \rightarrow (1-p)\rho + p \sigma \rho \sigma$, where $\sigma$ is
    a Pauli matrix specified by the user.
  \item \cinline{AmplitudeDampingChannel}: A single qubit channel with Kraus
    operators $|0\rangle \langle 0| + \sqrt{1-\gamma}|1\rangle \langle 1|$,
    $\sqrt{\gamma}|0\rangle \langle 1|$ . The user can additionally specify a
    non-zero excitation rate (for $|0\rangle \rightarrow |1\rangle$), i.e.
    implementing a generalized amplitude damping channel.
\end{itemize}
Each of the ``Global" channels also supports a corresponding ``Local'' version,
with noise applied independently to each qubit specified. The user can specify
unique parameters for each qubit, or a single set of values to use on every
qubit. Moreover, these noise channels can be `attached' to standard
\cinline{gate.Gate}, in order to implement a noisy gate.

\subsection{Grover Oracle}
In this section, we demonstrate how simple it is to construct a Grover gate
\cite{grover1996fast} in \HybridQ. The first step consists in providing a function
\cinline{f} to update the quantum state:
\begin{lstlisting}[language=Python]
from hybridq.gate import Projection
import numpy as np

def grover(self, psi, order):
    """
    Given a quantum state `psi` with qubits given 
    by `order`, invert the phase of the state
    0....0.
    """
    
    # First, let's create a projector
    proj = Projection(state='0' * self.n_qubits, 
                      qubits=self.qubits)
    
    # Project quantum state
    psi0, new_order = proj(psi=psi, 
                           order=order,
                           renormalize=False)
    
    # Check that order hasn't changed
    assert(new_order == order)
    
    # Update quantum state
    psi -= 2 * psi0
    
    # Return quantum state
    return psi, order
\end{lstlisting}
Once the Grover oracle is defined, a \cinline{FunctionalGate} can be used to
define the Grover gate:
\begin{lstlisting}[language=Python]
from hybridq.gate import FunctionalGate
import numpy as np

# Define a Grover gate that will act
# on qubits 1 and 2 only
grover_gate = FunctionalGate(f=grover, 
                             qubits=[1, 2])

# Define a superposition of three qubits
psi = np.ones(shape=(2, 2, 2), 
              dtype='complex')
psi /= np.linalg.norm(psi)

# Apply Grover gate to state
new_psi, _ = grover_gate(psi, order=[0, 1, 2])

# Print output
for i, x in enumerate(new_psi.ravel()):
    print('{0}: {1:+g}'.format(
      bin(i)[2:].zfill(3), 
      np.real_if_close(x)))
> 000: -0.353553
> 001: +0.353553
> 010: +0.353553
> 011: +0.353553
> 100: -0.353553
> 101: +0.353553
> 110: +0.353553
> 111: +0.353553
\end{lstlisting}

\section{Circuits}

If gates are the ``instructions`` to manipulate quantum states, circuits are
nothing less than ``programs'' with multiple instructions to follow. Following
the same philosophy we used to build gates, the circuit design in \HybridQ\ is
kept as simple as possible, with multiple ``utilities'' acting on them.

There are two classes of circuits in \HybridQ: regular circuits
acting on quantum states (\cinline{circuit.Circuit}) and the ``super'' circuits acting on
density matrices (\cinline{dm.circuit.Circuit}, with the latter being a generalization of the former) and
supports the inclusion of super operators. Indeed, while \cinline{circuit.Circuit}
can only accept \cinline{gate.Gate}, \cinline{dm.circuit.Circuit} can accept both
\cinline{gate.Gate} and \cinline{dm.gate.Gate}:
\begin{lstlisting}[language=Python]
from hybridq.gate import Gate
from hybridq.circuit import Circuit
from hybridq.dm.gate import Gate as SuperGate
from hybridq.dm.circuit import Circuit as SuperCircuit

# Generate a few gates
g1 = Gate('X')
g2 = Gate('PROJECTION', state='01')
g3 = SuperGate('KRAUS', gates=(Gate('RX'), Gate('RY')))

# Generate a regular circuit
Circuit([g1, g2])
> Circuit([
> 	Gate_X(name='X', 
>          n_qubits=1, 
>          M=numpy.ndarray(shape=(2, 2), 
>          dtype=int64)),
> 	ProjectionGate(name='PROJECTION', 
>                  n_qubits=2, 
>                  state='01')
> ])

# Generate a super circuit
SuperCircuit([g1, g2, g3])
> Circuit([
> 	Gate_X(name='X', 
>          n_qubits=1, 
>          M=numpy.ndarray(shape=(2, 2), 
>          dtype=int64)),
> 	ProjectionGate(name='PROJECTION', 
>                  n_qubits=2, 
>                  state='01')
> 	KrausSuperGate(name='KRAUS', 
>                  gates=(...), 
>                  s=1)
> ])

# Adding super gates to regular circuits fails
Circuit([g1, g2, g3])
> ValueError: 'KrausSuperGate' not supported. 
\end{lstlisting}

\subsection{Circuit utilities}

While regular and super circuits are kept simple, \HybridQ\ provides powerful
utilities to manipulate them. All circuit utilities can be found in
\cinline{circuit.utils} and \cinline{dm.circuit.utils}
respectively. Among them, it is worth mentioning:
\begin{itemize}
  \item \cinline{simplify(circuit)}: Simplify \cinline{circuit} as much as possible. If
    \cinline{use_matrix_commutation=True}, commutation using the matrix
    representation of gates is used to maximize the simplification,
  \item \cinline{matrix(circuit)}: Return the matrix representation of \cinline{circuit},
  \item \cinline{to_matrix_gate(circuit)}: Convert \cinline{circuit} to a
    \cinline{gate.MatrixGate},
  \item \cinline{isclose(a, b)}: Given two circuits \cinline{a} and \cinline{b},
    determine if the two circuits are close within an absolute tollerance of
    \cinline{atol}. If \cinline{use_matrix_commutation=True}, commutation using
    the matrix representation of gates is also used, 
  \item \cinline{compress(circuit)}: Compress gates in \cinline{circuit} so that
    gates in the new circuit will not have more than \cinline{max_n_qubits}
    qubits, Compression is very important to improve the performance of circuit
    simulations. Indeed, larger gates may better exploit vectorization
    instructions present in modern hardware and, therefore, reduce the
    computational cost to simulate the circuit,
  \item \cinline{pop(circuit, pinned_qubits)}: Remove gates in \cinline{circuit}
    outside the lightcone generated by \cinline{pinned_qubits}. Gates can be
    removed from either side of the circuits (\cinline{direction}).
\end{itemize}

\begin{lstlisting}[language=Python]
from hybridq.gate import Gate
from hybridq.circuit import Circuit, utils

# Generate a random circuit
circuit = Circuit(
    Gate('CPHASE', qubits=[q1, q2], params=[1]) 
      for q1 in range(5)
      for q2 in range(q1 + 1, 5))

# Compress gates
Circuit(map(utils.to_matrix_gate, 
            utils.compress(circuit, 
                           max_n_qubits=3)))
> Circuit([
> 	MatrixGate(name='MATRIX', qubits=(0, 1, 2), ...)
> 	MatrixGate(name='MATRIX', qubits=(0, 3, 4), ...)
> 	MatrixGate(name='MATRIX', qubits=(1, 3, 4), ...)
> 	MatrixGate(name='MATRIX', qubits=(2, 3, 4), ...)
> ])

# Shuffle circuit and invert it
inv_circuit = Circuit(circuit[x]
  for x in np.random.permutation(
    len(circuit))).inv()

# Simplify
utils.simplify(circuit + inv_circuit)
> Circuit([
> ])

# If use_matrix_commutation is False, circuit 
# and circuit_inv cannot cancel each other
assert (len(utils.simplify(circuit + inv_circuit,
            use_matrix_commutation=False)) == 0)
> AssertionError

\end{lstlisting}

\section{Simulations}

In the last few years, following the steady improvement of quantum technology to
build increasingly large quantum processors, different numerical algorithms have
been designed to reduce the computational cost to simulate quantum circuits.
With a few exceptions, the most successful among them can be classified in
one of the following categories: state vector simulation
\cite{guerreschi2020intel, pednault2019leveraging}, tensor contraction
\cite{villalonga2020establishing, villalonga2016flexible, gray2021hyper,
zlokapa2020boundaries, pan2020contracting, markov_tn, markov2018quantum}, and
expansion around efficiently simulable quantum circuits
\cite{aaronson_gottesmann_stabilizer_sim, Bravyi2019simulationofquantum,
bravyi2016improved, jozsa2008matchgates, aharonov2003simple}.

All of these techniques have pros and cons, and their performance largely
depends on the structure of the quantum circuit to simulate. For instance, the
computational cost of state vector simulations grows only polynomially with the
number of gates but its memory requirement increases exponentially with the
number of qubits, limiting it to $\sim$\,\!50 qubits \cite{markov2018quantum}.
On the other hand, tensor contraction \cite{markov_tn} with its “slicing”
technique \cite{markov_tn, markov2018quantum} allows the simulation of circuits
with a large number of qubits \cite{villalonga2016flexible,
villalonga2020establishing}.  However, its computational complexity is bounded
by the treewidth of the quantum circuit and it quickly grows (often
exponentially) with its depth. The expansion around efficiently simulable
circuits \cite{bravyi2016improved, jozsa2008matchgates, aharonov2003simple} may
allow simulations of deep quantum circuits with multiple qubits if the
expansion has a limited number of terms. One of the most well-known expansions
is around stabilizer circuits made of Clifford gates only
\cite{aaronson_gottesmann_stabilizer_sim, bravyi2016improved}. In this case,
the complexity grows exponentially with the number of non-Clifford gates
\cite{Bravyi2019simulationofquantum}.  While stabilizer circuits (which contain
only Clifford gates) are widely used in error-correction schemes,
hard-to-simulate circuits \cite{bravyi2016improved} and application-driven
circuits \cite{hastings2014improving, wecker2014gate} often have a large number
of non-Clifford gates, making this technique unsuitable for such cases.

In the spirit to keep the library easy to use and allow the development of
hybrid algorithms (giving rise to the name \HybridQ), all simulation techniques
in \HybridQ\ can be called using a single entry point, which uses the same
syntax, regardless of the choice of the numerical methods or hardware. 
More precisely, \HybridQ\ provides \cinline{circuit.simulation.simulate} and
\cinline{circuit.dm.simulation.simulate} to simulate quantum state and density
matrices respectively, with the latter entry point being a macro which transforms
a super circuit to a regular circuit and eventually calls the former entry
point. At the moment, \HybridQ\ provides support to simulate both
quantum state and density matrices using state vector simulation
(\cinline{optimize='evolution'}), tensor contraction (\cinline{optimize='tn'}),
and Clifford expansion (\cinline{optimize='clifford'}). 

\HybridQ\ fully supports HPC simulations for both \cinline{optimize='tn'} and
\cinline{optimize='clifford'}, but we are planning to include MPI
parallelization for \cinline{optimize='evolution'} in the next \HybridQ\
release. To parallelize \HybridQ\ simulations on HPC clusters, it suffices to
call the Python script using \cinline{mpiexec} from any terminal: \HybridQ\ will
automatically detect the use of \texttt{MPI} and split the workload among
different nodes.

\subsection{Initial and Final State Specification}

In \HybridQ, the initial and final quantum states are specified by using the
keywords \cinline{initial_state} and \cinline{final_state}, 
that is, the final
quantum state is projected to the provided state (for \cinline{optimize='tn'}
only). In both cases, the quantum states can be provided by using either arrays,
strings, or \cinline{circuit.Circuit} (for the simulation of density matrices only).
Arrays must be \cinline{numpy.ndarray} convertible, with a number of dimensions
proportional to the number of qubits (for quantum state simulations) or twice
the number of qubits (for density matrix simulations). At the moment, \HybridQ\
only supports two-state (qubit) dimensions, but we are planning to allow $d$-state
(``qudit") dimensions with $d \geq 2$ in the upcoming release. 

Similarly, strings must have either $n$ chars (for quantum state
simulations) or $2n$ chars (for density matrix simulations), with $n$ being the
number of qubits. The allowed tokens are:
\begin{itemize}
  \item \cinline{0}, \cinline{1}: Set qubit to either \cinline{0} or
    \cinline{1} state in the computational basis,
  \item \cinline{+}, \cinline{-}: Set qubit to either \cinline{+} or
    \cinline{-} state in the X basis,
  \item \cinline{.}: The corresponding qubit is left ``open'' while
    contracting the circuit (for \cinline{optimize='tn'} only),
  \item \cinline{[a-zA-Z]}: If two or more qubits have the same letter, a
    multi-index Kronecker delta is attached to them (that is, qubits are
    traced-out together. For \cinline{optimize='tn'} only). This is particularly
    useful to trace out subsets of qubits while performing the simulation of
    density matrices.
\end{itemize}

\subsection{State Vector}

\HybridQ\ provides state vector simulation of the quantum state with the keyword
\cinline{optimize='evolution'}. Currently, \HybridQ\ supports two different
numerical engines to perform matrix-vector multiplications:
\cinline{optimize='evolution-hybridq'} (the default value), which uses a
in-house developed C$^{++}$ core that exploits AVX instructions and
parallelization via \texttt{OpenMP}, and \cinline{optimize='evolution-einsum'},
which uses \texttt{NumPy} \cite{numpy} and \texttt{opt-einsum} \cite{opt_einsum}
as the numerical backend. 

For the C$^{++}$ backend, state vectors are stored as a pair of contiguous array
of AVX packed floating numbers ($8$-\cinline{float}/$4$-\cinline{double} for
\texttt{AVX2}), one for the real part and one for the imaginary of the state
vector. If the matrix-vector multiplication involves the least significative
qubits, qubits are swapped to fully exploit AVX instructions. For the
\cinline{einsum} backend, state vectors are stored as \cinline{numpy.ndarray} of
complex numbers.

Numerical simulations on GPUs and TPUs are enabled for
\cinline{optimize='evolution-einsum'} by using \texttt{JAX} \cite{jax} as the backend
(\cinline{backend='jax'}). 
Multi-threaded optimization is supported for
\cinline{optimize='evolution-hybridq'} only via \texttt{OpenMP}, and the number
of used threads can be tuned by modifying the environment variable
\cinline{OMP_NUM_THREADS} (if not set, all cores are used by default). At the
moment, \cinline{optimize='evolution'} does not support HPC simulations using
\texttt{MPI}, but we are planning to include such feature in the upcoming
release.

\begin{lstlisting}[language=Python]
from hybridq.dm.circuit.simulation import simulate \
  as dm_simulate
from hybridq.circuit.simulation import simulate
from hybridq.extras.random import get_rqc
from hybridq.circuit import utils
from tqdm.auto import tqdm
from time import time
from os import system
import numpy as np

# Get CPU name
system('lscpu | egrep "Model name|^CPU\(s\)"')
> CPU(s):     8
> Model name: Intel(R) Core(TM) i7-8550U CPU @ 1.80GHz

# Get random circuit
circuit = get_rqc(n_qubits=6, n_gates=50)

# Set superposition as initial state
initial_state = '+' * 6

# Simulate circuit using einsum on CPU
psi_cpu = simulate(circuit,
                   initial_state=initial_state,
                   optimize='evolution-einsum')

# Simulate circuit using einsum on GPU
psi_gpu = simulate(circuit,
                   initial_state=initial_state,
                   optimize='evolution-einsum',
                   backend='jax')


# Simulate density matrix using C++ core on CPU
dm_cpu = dm_simulate(circuit,
                     initial_state=2*initial_state,
                     optimize='evolution')

# Get density matrix from quantum state
dm_psi = np.reshape(np.kron(psi_cpu.ravel(),
                            psi_cpu.ravel().conj()), 
                    (2, ) * 12)

# Checks
assert (np.allclose(psi_cpu, psi_gpu))
assert (np.allclose(dm_psi, dm_cpu))

# Get some random circuits
cs = {n: [get_rqc(n, n**2) for _ in range(5)] 
        for n in tqdm(range(12, 29, 2), 
        desc='Generating random circuits')}

# Compute typical depth
depth = {n:np.average([len(utils.moments(c)) 
          for c in cs[n]]) for n in cs}
print('Typical depth')
print('-------------')
for n,d in depth.items():
    bar = '@' * int(d / 10)
    print(f'{n}: {bar:s} {d:1.2f}')
> Typical depth
> -------------
> 12: @@@ 32.82
> 14: @@@@ 42.82
> 16: @@@@ 49.18
> 18: @@@@@ 52.73
> 20: @@@@@@ 61.36
> 22: @@@@@@ 68.82
> 24: @@@@@@@ 73.73
> 26: @@@@@@@@ 84.00
> 28: @@@@@@@@@ 90.00

# Initialize times
times = {n: [] for n in cs}

# Simulate circuits
for n in tqdm(range(12, 29, 2), 'Simulating'):
    for c in tqdm(cs[n], leave=False):
        t_ini = time()
        psi = simulate(c,
                 initial_state='+',
                 optimize='evolution',
                 max_largest_intermediate=2**30)
        t_end = time()
        times[n].append(t_end - t_ini)

print('Typical runtime (s)')
print('-------------------')
for n,t in [(n,np.average(times[n])) 
              for n in times]:
    bar = '@' * max(1, int(t / 2))
    print(f'{n}: {bar:s} {t:1.2f}')
> Typical runtime (s)
> -------------------
> 12: @ 0.83
> 14: @ 1.12
> 16: @ 1.49
> 18: @ 1.98
> 20: @ 2.52
> 22: @ 3.39
> 24: @@ 5.22
> 26: @@@@@ 11.50
> 28: @@@@@@@@@@@@@@@@@@@ 38.81
\end{lstlisting}

\subsection{Tensor Contraction}

Tensor contraction \cite{markov_tn} has been shown to be a particularly
powerful approach to simulating quantum circuits in the Noisy
Intermediate-Scale Quantum regime \cite{preskill2018quantum, q_suprem_google,
villalonga2020establishing, villalonga2016flexible, mi2021information,
gray2021hyper, huang2020classical, pan2020contracting}. \HybridQ\ fully
supports the simulation of quantum circuits using tensor contraction via
\texttt{Quimb} \cite{quimb} and \texttt{CoTenGra} \cite{cotengra} More precisely, given
a \cinline{Circuit}, and both the \cinline{initial_state} and
\cinline{final_state}, \HybridQ\ automatically identifies the best contraction
(using \texttt{CoTenGra}), and eventually performs the tensor network
contraction (using \texttt{Quimb}) to simulate the quantum circuit. If the
largest intermediate tensor is larger than the provided
\cinline{max_largest_intermediate}, slicing is automatically applied to the
tensor (thanks to \texttt{CoTenGra}). Multiple parameters can be passed to both
\texttt{Quimb} and \texttt{CoTenGra} through the
\cinline{circuit.simulation.simulate} entrypoint (use the \cinline{help}
command to check all the supported options).
Depending on the backend, multi-thread optimization is supported via either
\texttt{OpenMP} or \texttt{MKL}, and the number of used threads can be tuned by
modifying the environment variable \cinline{OMP_NUM_THREADS} and
\cinline{MKL_NUM_THREADS} respectively. Finding the optimal contraction scheme
can be also parallelized, which is activated by using \cinline{parallel=True}
(\cinline{False} by default).

\HybridQ\ fully supports \texttt{MPI} parallelization for \cinline{optimize='tn'}.
More precisely, slices are automatically distributed among the available nodes.
Once completed, slices are then recollected using a divide-and-conquer approach
to avoid overloading a single node. 

\begin{lstlisting}[language=Python]
from hybridq.dm.circuit.simulation import simulate \
  as dm_simulate
from hybridq.circuit.simulation import simulate
from hybridq.extras.random import get_rqc
import numpy as np

# Get random circuit
circuit = get_rqc(n_qubits=6, n_gates=50)

# Set superposition as initial state
initial_state = '+' * 6

# Project all qubits to the state +, 
# excluding qubit 1 which is left "open"
final_state = '+.++++'

# Simulate using tensor contraction (on GPU)
psi = simulate(circuit,
               initial_state=initial_state,
               final_state=final_state,
               optimize='tn', parallel=True,
               backend='jax')

# For the density matrix simulation, let's add
# some extra qubits to eventually trace out
dm_circuit = circuit + get_rqc(2, 5, indexes=[6, 7])

# Let's trace out qubits 6 and 7
dm_initial_state = 2 * (initial_state + '++')
dm_final_state = 2 * (final_state + 'ab')

# Simulate density matrix using tensor contraction
dm = dm_simulate(dm_circuit,
                 initial_state=dm_initial_state,
                 final_state=dm_final_state,
                 parallel=True, optimize='tn')

# Get density matrix from quantum state
dm_psi = np.reshape(np.kron(psi.ravel(),
                            psi.ravel().conj()), 
                    (2, ) * 2)

# Check
assert (np.allclose(dm_psi, dm))
\end{lstlisting}

\subsection{Clifford Expansion}

As demonstrated by Gottesman and Knill \cite{gottesman1998heisenberg}, circuits made of Clifford gates
only (i.e., gates which are stabilizers of the Pauli group) can be simulated in
polynomial time \cite{aaronson_gottesmann_stabilizer_sim}. Such circuits are called ``stabilizer'' circuits.
Similarly, a ``stabilizer state" is any quantum state that can be obtained by applying a stabilizer
circuit to a state in the computational basis. Such states are widely used in quantum error-correction. More generally, if a
circuit is made of $k$ non-Clifford gates, the simulation cost is exponential in
$k$, with a power which depends on the non-Clifford gate. The exponential cost
derives from the ``expansion'' of the non-Clifford gates,
which eventually requires the simulation of $\chi$ independent stabilizer
circuits, with $\chi$ called the stabilizer ``rank'' \cite{aaronson_gottesmann_stabilizer_sim, bravyi2016improved}. For instance, controlled
phases can be expanded using $\chi=2$ stabilizers:
\begin{align*}
  \text{CZ}(\alpha) &= \mathbb{I} + (e^\alpha - 1)
  \left|11\right\rangle\!\!\left\langle11\right|,
\end{align*}
where we used the fact that qubits in stabilizer states can always be projected to 
either $0$ or $1$ in the computational basis in polynomial time. In general, if a
$k-$qubit non-Clifford gate is applied to a stabilizer state, the resulting
non-stabilizer state can always be expanded 
using no more than
$\chi = 2^p \leq 2^k$
stabilizers, with $p$ being the number of elements different from zero in the
matrix representation of the non-Clifford gate.

To simulate density matrices in \HybridQ\ (\cinline{optimize='tn'}), we start
from a Pauli ``string'' of the form $P = P_0 \otimes P_1 \otimes\,\ldots\,\otimes
P_{n-1}$, with $n$ being the number of qubits, and gates $\mathcal{G}_i$ are applied
one by one as $\mathcal{G}\,P^\prime\,\mathcal{G}^\dagger$.
If $\mathcal{G}$ is a Clifford gate, since it stabilizes the Pauli group, $P^\prime$ will just be a
different Pauli string (that is, Clifford gates ``move'' Pauli strings to Pauli
strings). However, if $\mathcal{G}$ is a non-Clifford gate, $P^\prime$ will be
expanded as superposition of orthogonal Pauli strings \cite{mi2021information}:
$$
  P^\prime = \sum_{i=1}^\chi \omega_i P^i_0 \otimes P^i_1
  \otimes\,\ldots\,\otimes P^i_{n-1}.
$$
That is, the simulation of the density matrix ``branches'' $\chi$ times. Once
all the gates are applied, the final density matrix will be a superposition of
orthogonal Pauli strings. Observe that the amount of memory required to store a
density matrix expanded in Pauli string is $\mathcal{O}(n\,\chi)$, with $\chi$
being the dominant term. If few non-Clifford gates are applied, all the relevant
details of the density matrix can be stored in memory \cite{mi2021information}.\\

To further improve the performance of simulating density matrices by Clifford
expansion, we introduce in \HybridQ\ 
the two following innovations:
\begin{itemize}
  \item Rather than applying gates one by one, \HybridQ\ compresses multiple
    gates into larger ones, which are eventually applied to the Pauli strings.
    This approach drastically reduces the amount of branches by avoiding
    unnecessary branches (see example below). Indeed, consider the simple
    example of multiple non-Clifford gates applied to the same group of qubits.
    If each gate $\mathcal{G}_i$ were applied one by one, the number of
    branches would be the product of each single $\chi = \prod_i \chi_i$.
    However, if compressed together to a single $k-$qubit gate, the total number
    of branches may only be $\chi \leq 4^k$. 
  \item Instead of expanding all the non-Clifford gates at the beginning of the
    simulation, we dynamically branch the Pauli strings by using a depth-first
    expansion. While this approach may require more memory than expanding all
    the non-Clifford gates beforehand, the dynamical branching effectively
    reduces the amount of branches \cite{mi2021information} to simulate, consequently reducing
    the computational cost to simulate the density matrix. Also, the depth-first
    expansion ensures that the amount of memory required to keep track of
    branches grows only polynomially with the number of non-Clifford gates.
\end{itemize}

To achieve compile-time performance, the core parts of the Clifford expansion
are Just-In-Time (JIT) compiled using \texttt{Numba} (therefore, the first time
\cinline{optimize='clifford'} is used, it may take a while).

\HybridQ\ fully supports multi-thread parallelization for
\cinline{optimize='clifford'}, which is enabled by using
\cinline{parallel=True}. When enabled, an initial breadth-first expansion is
used to gather multiple branches which are eventually distributed among
different threads. HPC simulations via \texttt{MPI} are also fully supported for
\cinline{optimize='clifford'}. Similarly, a breadth-first expansion to gather
multiple branches which are eventually distributed to multiple nodes.
\texttt{MPI} and \cinline{parallel=True} can be used together to maximize the
performance.

\begin{lstlisting}[language=Python]
from hybridq.gate import Gate
from hybridq.dm.circuit.simulation import simulate \
  as dm_simulate
from hybridq.extras.random import get_rqc
from hybridq.circuit import utils
from hybridq.utils import kron
import numpy as np

# Get random circuit
circuit = get_rqc(6, 20)

# Get random paulis
paulis = [
    Gate(np.random.choice(list('IXYZ')), [q]) 
      for q in circuit.all_qubits()
]

# Simulate density using clifford expansion
# (no compression)
pauli_dm_0, info_0 = dm_simulate(circuit,
                           initial_state=paulis,
                           compress=0,
                           parallel=True,
                           optimize='clifford',
                           return_info=True)

# Simulate density using clifford expansion
# (with compression)
pauli_dm_4, info_4 = dm_simulate(circuit,
                           initial_state=paulis,
                           compress=4,
                           parallel=True,
                           optimize='clifford',
                           return_info=True)

for c,info in zip((0, 4), (info_0, info_4)):
    n = info['n_explored_branches']
    r = info['runtime (s)']
    print(f'Explored {n:7,} branches in', end=' ') 
    print(f'{r:1.3f} seconds (compress={c})')
> Explored 745,614 branches in 2.530s (compress=0)
> Explored  19,273 branches in 0.425s (compress=4)

# Reconstruct density matrices
dm_0 = sum(w * kron(*[Gate(g).matrix() for g in p]) 
            for p,w in pauli_dm_0.items())
dm_4 = sum(w * kron(*[Gate(g).matrix() for g in p]) 
            for p,w in pauli_dm_4.items())

# Check reconstruction
assert (np.allclose(dm_0, dm_4, atol=1e-5))

# Compute density matrix directly from 
# 'circuit' and 'paulis'
dm_full = utils.matrix(circuit + \
                       paulis + \
                       circuit.inv())

# Check
assert (np.allclose(dm_0, dm_full, atol=1e-5))

\end{lstlisting}

\section{Simulations with noise}

\HybridQ\ supports the inclusion of noise at the Kraus operator level (see
Section~\ref{sec:noisy_gates}). In particular, it is possible to implement generic
transformations of the form 
\begin{equation}
    \Lambda[X] = \sum_{i,j} s_{ij} L_i X R_j^\dag,
\end{equation}
thus allowing in principle \textit{any} quantum map to be represented
\cite{q_computation_nielsen_chuang}.

\begin{figure}
    \centering
    \includegraphics[width=0.98\columnwidth]{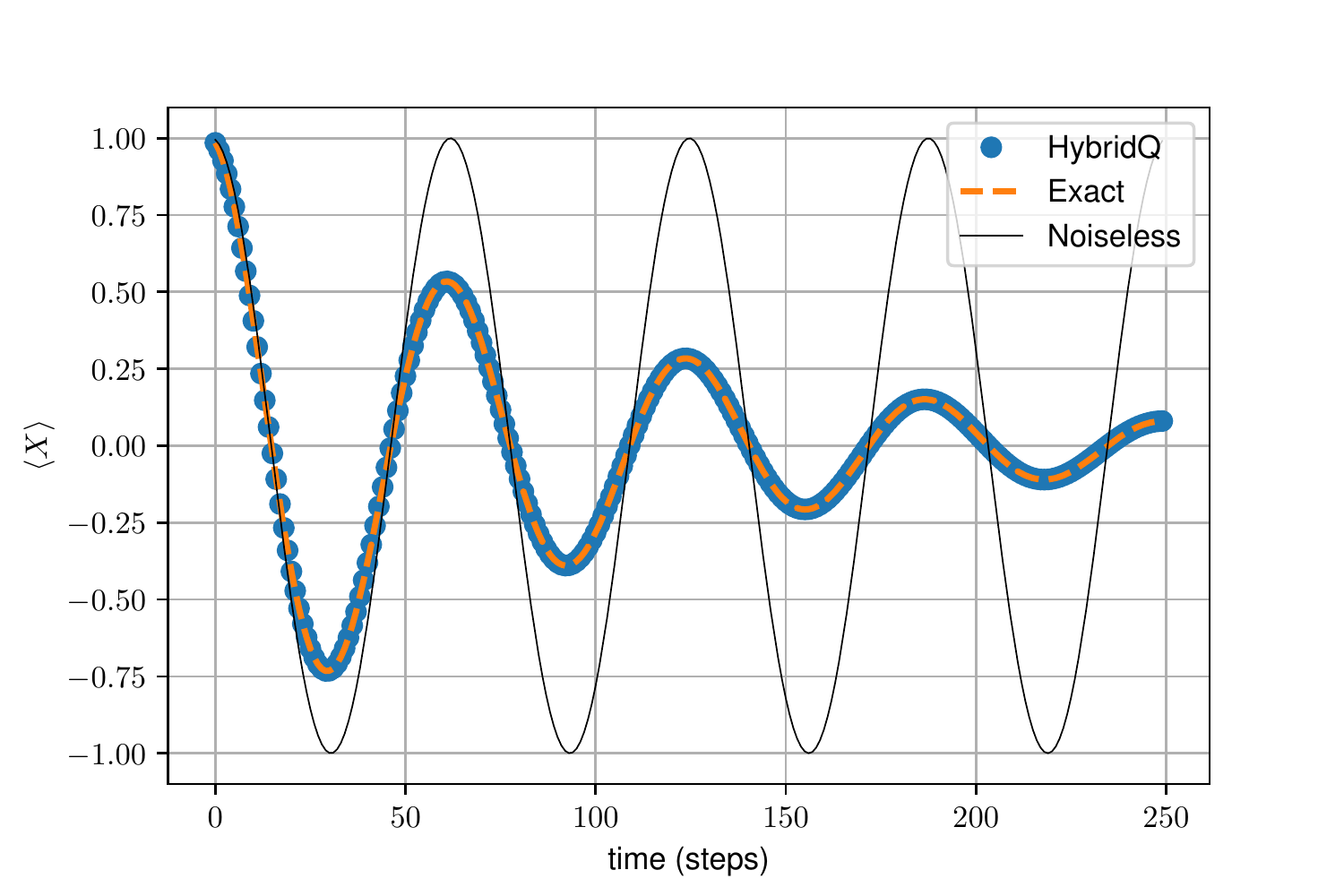}
    \caption{Depolarizing noise during a $Z$ rotation on a qubit starting in
    the $|+\rangle$ state. Plot shows the expectation value along $X$. Each
    step is a rotation by a fixed amount.}
    \label{fig:depol_1q}
\end{figure}

At the highest level of generality, a quantum map can be specified by the left
$(L_i)$ and right $(R_j)$ operators, the matrix of coefficients $s_{ij}$, as
well as the qubits on which the noise acts.  A user may also define a quantum
map much more concisely through an alternative API, or use the pre-defined noise
models as in Section~\ref{sec:noisy_gates}.  An example of a single qubit
\cinline{DepolarizingChannel} during a $Z$ rotation is shown in
Fig.~\ref{fig:depol_1q}.

\HybridQ\ also allows noise models to be easily built on top of pre-existing
circuits. For example, one can attach two-qubit noise channels to the two-qubit
gates in a circuit. We have a code snippet below which shows such an example.

Noise can be simulated either at the density matrix level, or by pure state
sampling of the Kraus operators. Whilst the former is exact, the system reachable sizes
more quickly become prohibitive since the density matrix scales
as $4^n$ for $n$ qubits (instead of $2^n$ for a pure state). Pure state sampling allows one to
reach larger system sizes than is possible through density matrix simulation
alone, at the expense of accuracy (from the sampling). The code is designed to
support seamlessly change between the two paradigms by setting a single argument.
Thus, the user can perform full density matrix simulations up to the size they
can reach within the limits of their hardware, and then switch to sampling
mode. 

\subsection{Performance}

Noisy simulations are fully compatible with all of the backends available for
pure state simulation. This enables one to perform noisy
simulations using the $\text{C}^{++}$ backend, GPU's, or even via tensor network
contraction. Here, we compare some wall-clock speed tests for \HybridQ\ versus
one of the most well-known simulators of open quantum systems, QuTiP (version
4.7) \cite{qutip2}. All tests are performed on the same machine with a 2.5 GHz
Quad-Core Intel Core i7 chip, and 16GB RAM%
\footnote{We use the latest version
of QuTiP as of this writing, installed from source with OpenMP, with system
specifications: QuTiP Version: 4.7.0.dev0+nogit,
Numpy Version: 1.19.2,
Scipy Version: 1.5.0,
Cython Version: 0.29.24,
Matplotlib Version: 3.2.2,
Python Version: 3.8.5,
Number of CPUs: 4,
BLAS Info: INTEL MKL,
OPENMP Installed: True,
INTEL MKL Ext: True}.
All notebooks to reproduce results in this Section are available at \cite{GH_HybridQ}.

\begin{figure}
    \centering
    \includegraphics[width=0.98\columnwidth]{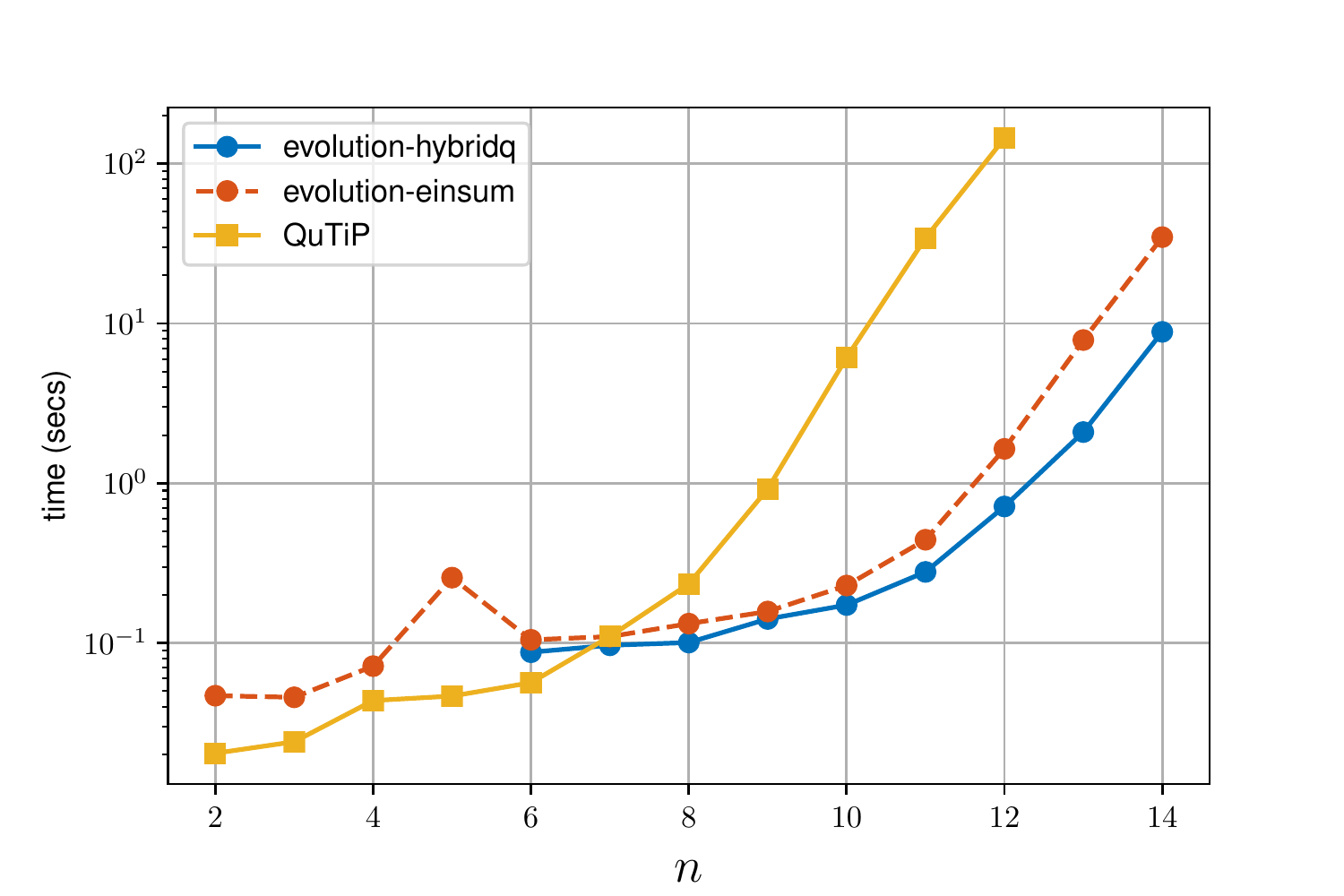}
    \caption{Comparison of QuTiP to \HybridQ\ speed on density matrix
    operations. We show the wall-clock time versus system size $n$ (number of
    qubits). We apply a depolarizing channel on each qubit of a random density matrix,
    utilising two of the \HybridQ\ backends. Note the `evolution-hybridq` backend only works for more than 5 qubits in the density matrix setting.}
    \label{fig:depol_test}
\end{figure}

In Fig.~\ref{fig:depol_test} we show the time taken to apply a local
depolarizing channel on each qubit in the system. Shown by the circular markers
are two \HybridQ\ protocols, one using \texttt{NumPy}'s \texttt{einsum} (\cinline{optimize='evolution-einsum'}, red-dash), and the
other the $\text{C}^{++}$ core (\cinline{optimize='evolution-hybridq'}, blue-solid). We find after around $7$
qubits, \HybridQ\ 
substrantially outperforms QuTiP. This is likely due to the fact
that QuTiP represents objects in a sparse format which means we generally
expect the simulations will be slower, especially at larger sizes. 
In the comparison we also include all the time to set-up and prepare the circuit
for the simulation in \HybridQ\ which involves some pre-processing, thus there
is likely a greater overhead at the smaller system sizes.

\begin{figure}
    \centering
    \includegraphics[width=0.98\columnwidth]{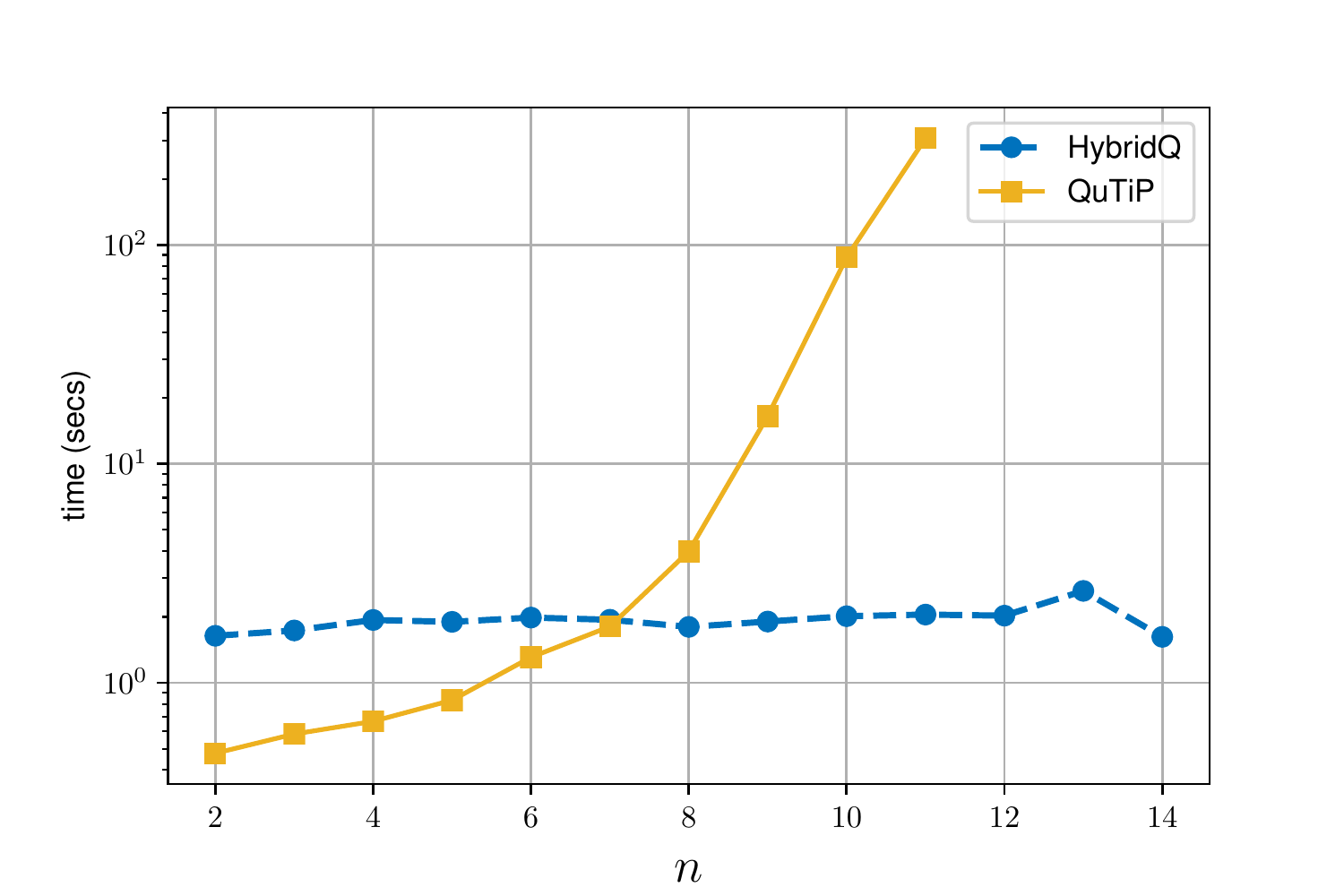}
    \caption{Tensor network contraction on a density matrix circuit. For each
    system size, we generate a random quantum circuit of $50$ total gates, which
    we then add depolarizing noise after each gate. Using the tensor network
    contraction, we extract the reduced state of the 0'th qubit. In QuTiP we
    obtain this state by performing the full simulation and tracing out the
    other degrees of freedom.}
    \label{fig:tn_test}
\end{figure}

Lastly, in Fig.~\ref{fig:tn_test} we perform a particular simulation using the
built-in tensor network contraction. Here we find an even starker performance
difference. In this simulation, we first construct a random quantum circuit.
Then, after each \cinline{Gate}, we add a \cinline{GlobalDepolarizingChannel} with
$1\%$ noise (acting on one or both qubits of the gate respectively). We consider
the task of extracting the reduced density matrix of one particular qubit. This
can be achieved via the \cinline{optimize='tn'} optimization method in \HybridQ\, which when
applied to a density matrix is equivalent to the partial trace (when we sum
over the left and right indices of the density matrix as required).  Some
example code to find the reduced density matrix of qubit $0$ via tensor contraction, in a noisy
simulation, with initial state being the $+$ state, is shown below:

\begin{lstlisting}[language=Python]
from hybridq.dm.circuit.simulation import simulate as dm_simulate
from hybridq.extras.random import get_rqc
from hybridq.noise.utils import add_depolarizing_noise

# Generate a random quantum circuit
circuit = get_rqc(n_qubits=3, n_gates=200)

# Add 1% noise to single qubit gates, 
# and 1.5% to two qubit gates
noisy_circuit = add_depolarizing_noise(circuit, 
                                       probs=(0.01, 
                                       0.015))

# Simulate by contracting qubit 1,2 indices,
# and leave 0 open
rho_0 = dm_simulate(noisy_circuit, 
                    initial_state='+', 
                    final_state='.ab.ab', 
                    optimize='tn')

\end{lstlisting}

We find that \HybridQ\ scales very favorably here (note that the number of
gates is not scaling with $n$). In contrast, there is no native way to perform
such a calculation in QuTiP (as far as we are aware). We therefore perform the
entire state evolution, and trace out the unwanted qubits in that case. 

\section{Summary}

In this paper we presented \HybridQ, a highly  extensible  platform designed to
provide a common framework to integrate multiple state-of-the-art techniques to
run on a variety of hardware.  \HybridQ\ provides a simple and expressive
language that allows seamless switching from one technique to another as well as
from one hardware to the next, without the need to write lengthy translations,
thus greatly simplifying the development of new hybrid algorithms and
techniques. 

At the moment, \HybridQ\ can simulate quantum circuits using state-vector
evolution, tensor contraction and Clifford expansion. All the aforementioned
methods allow multi-thread parallelization on single nodes, while tensor
contraction and Clifford expansion allow HPC parallelization on multiple nodes
via \texttt{MPI} (state-vector parallelization on multiple nodes is planned on
the next \HybridQ\ release).

\HybridQ\ fully integrates noise simulations for all the supported simulation
techniques, by ``converting'' noisy circuits to regular circuits. For instance,
it is possible to compute expectation values of local operators of noisy
circuits by using tensor contraction, without the need to simulate the full
density matrix. This immediately translate to a faster simulation as shown in
Fig.~\ref{fig:tn_test}.

\section*{Code Availability}
\HybridQ\ is an open source tool and freely available under the 
\texttt{Apache 2.0} licence at
\href{https://github.com/nasa/hybridq}{GitHub@NASA}\cite{GH_HybridQ}.

\section*{Acknowledgments}
We are grateful for support from NASA Ames Research Center, particularly the
NASA Transformative Aeronautic Concepts Program, and also from DARPA under IAA
8839, Annex 114.  The authors acknowledge the support from the NASA Ames
Research Center and the support from the NASA Advanced Supercomputing Division
for providing access to the NASA HPC systems, Pleiades and Merope.  J.M. is
supported by NASA Academic Mission Services (NAMS), contract number NNA16BD14C.
SM. is supported by the Prime Contract No.  80ARC020D0010 with the NASA Ames
Research Center.
The United States Government retains, and by accepting the article for
publication, the publisher acknowledges that the United States Government
retains, a non-exclusive, paid-up, irrevocable, worldwide license to publish or
reproduce the published form of this work, or allow others to do so, for United
States Government purposes.

\appendix
\section{Artifact Description Appendix: HybridQ: A Hybrid Simulator for Quantum Circuits}

\subsection{Abstract}

\HybridQ\ is a highly extensible platform designed to provide a common
framework to integrate multiple state-of-the-art techniques to simulate large
scale quantum circuits on a variety of hardware. 

\subsection{Description}

The philosophy behind development of \HybridQ\ has been driven by
three main pillars: \emph{Easy to Use}, \emph{Easy to Extend}, and \emph{Use
the Best Available Technology}. The powerful tools of \HybridQ\ allow users
to manipulate, develop, and extend noiseless and noisy quantum circuits for different
hardware architectures. \HybridQ\ supports large-scale high-performance
computing (HPC) simulations, automatically balancing  workload among
different processor nodes and enabling the use of multiple backends to
maximize parallel efficiency.  Everything is then glued together by a simple
and expressive language that allows seamless switching from one technique to
another as well as from one hardware to the next, without the need to write
lengthy translations, thus greatly simplifying the development of new hybrid
algorithms and techniques.

\subsubsection{Software installation}

\HybridQ\ can be installed by either using \texttt{pip}:
\begin{lstlisting}[language=bash]
pip install git+https://github.com/nasa/hybridq
\end{lstlisting}
or by using \texttt{conda}:
\begin{lstlisting}[language=bash]
conda env create -f envinronment.yml
\end{lstlisting}

\subsubsection{Software dependencies}

\HybridQ\ requires:
\begin{itemize}
  \item Python 3.7+
  \item C$^{++}$ compiler (optional)
\end{itemize}

All the other dependencies are installed through \texttt{pip} (see
\cite{GH_HybridQ} for the full list of requirements).

\subsection{Experiment workflow}

All scripts included in the main text are complete and reproducible.

\subsection{Evaluation and expected result}

All results are matched with either theoretical results or against well
established third-parties tools.

\subsection{Code Availability}
\HybridQ\ is an open source tool and freely available under the 
\texttt{Apache 2.0} licence at
\href{https://github.com/nasa/hybridq}{GitHub@NASA}\cite{GH_HybridQ}.

\IEEEtriggeratref{15}
\bibliographystyle{ieeetr}
\bibliography{refs}

\end{document}